\documentclass[prl,twocolumn,showpacs,nofootinbib,
superscriptaddress,preprintnumbers,amsmath,amssymb]{revtex4}
\usepackage{graphicx}
\usepackage{amsbsy}
\usepackage{dcolumn}
\usepackage{bm}
\begin{document}
  \title{Absence of broken time reversal symmetry 
    beneath the surface of \YBCO}
  \author{H. Saadaoui}
\email{hassan.saadaoui@psi.ch}
\affiliation{Laboratory for Muon Spin Spectroscopy, Paul Scherrer Institut,
  5232 Villigen PSI, Switzerland.}
\author{Z. Salman}
\affiliation{Laboratory for Muon Spin Spectroscopy, Paul Scherrer Institut,
  5232 Villigen PSI, Switzerland.}
\author{T. Prokscha}
\affiliation{Laboratory for Muon Spin Spectroscopy, Paul Scherrer Institut,
  5232 Villigen PSI, Switzerland.}
\author{A. Suter}
\affiliation{Laboratory for Muon Spin Spectroscopy, Paul Scherrer Institut,
  5232 Villigen PSI, Switzerland.}
\author{H. Huhtinen}
\affiliation{Wihuri Physical Laboratory, Department of Physics and Astronomy,
  University of Turku, FI-20014 Turku, Finland}
\author{P. Paturi}
\affiliation{Wihuri Physical Laboratory, Department of Physics and Astronomy,
  University of Turku, FI-20014 Turku, Finland}
\author{E. Morenzoni}
\affiliation{Laboratory for Muon Spin Spectroscopy, Paul Scherrer Institut,
  5232 Villigen PSI, Switzerland.}
\date{\today}

\newcommand{\LSCO}{La$_{2-x}$Sr$_x$CuO$_{4-\delta}$}
\newcommand{\YBCO}{YBa$_{2}$Cu$_{3}$O$_{7-\delta}$}
\newcommand{\BISCO}{Bi$_2$Sr$_2$Cu$_2$O$_{8+\delta}$}
\newcommand{\STO}{SrTiO$_{3}$}
\newcommand{\SRO}{Sr$_2$RuO$_{4}$}
\newcommand{\PCCO}{Pr$_{2-x}$Ce$_x$CuO$_{4-\delta}$}
\newcommand{\CO}{$\rm CuO_2$}
\newcommand{\Li}{${}^8$Li$^+$}
\newcommand{\Lip}{${}^8$Li$^+$}
\newcommand{\NbSe}{NbSe$_2$}

\newcommand{\msr}{$\mu$SR}
\newcommand{\lem}{LE-$\mu$SR}
\newcommand{\bnmr}{$\beta$-NMR}
\newcommand{\spct}{superconductivity}
\newcommand{\Spct}{Superconductivity}
\newcommand{\spc}{superconductor}
\newcommand{\Spc}{Superconductor}
\newcommand{\etal}{{\it et al.}}
\newcommand{\Htsc}{High-$T_c$ superconductors}
\newcommand{\htsc}{high-$T_c$ superconductors}
\newcommand{\ie}{{\it i.e.}}
\newcommand{\Tc}{$T_c$}

\newcommand{\PRL}[3]{Phys. Rev. Lett. {\bf #1}, {#2} ({#3})}
\newcommand{\PRB}[3]{Phys. Rev. B {\bf {#1}}, {#2} ({#3})}
\newcommand{\PB}[3]{Physica B {\bf {#1}}, {#2} ({#3})}
\newcommand{\PC}[3]{Physica C {\bf {#1}}, {#2} ({#3})}
\newcommand{\Nt}[3]{Nature {\bf {#1}}, {#2} ({#3})}
\newcommand{\Sc}[3]{Science {\bf {#1}}, {#2} ({#3})}
\newcommand{\RMP}[3]{Rev. Mod. Phys. {\bf {#1}}, {#2} ({#3})}
\newcommand{\JPSJ}[3]{J. Phys. Soc. Jap. {\bf #1}, {#2} ({#3})}
\newcommand{\RPP}[3]{Rep. Prog. Phys. {\bf {#1}}, {#2} ({#3})}
\newcommand{\ibid}[3]{{\it ibid}. {\bf {#1}}, {#2} ({#3})}
 \newcommand{\equ}[2]{\begin{equation}\label{#1}{#2}\end{equation}}
\newcommand{\meq}[2]{\begin{eqnarray}\label{#1}{#2}\end{eqnarray}}
\newcommand{\bq}{{\bf q}}
\newcommand{\bk}{{\bf k}}
\newcommand{\bkp}{{\bf k'}}
\newcommand{\br}{{\bf r}}
\newcommand{\bR}{\bf R}
\newcommand{\bp}{{\bf p}}

\begin{abstract}
  We report the results of a search for spontaneous magnetism
  due to a time reversal symmetry breaking phase
  in the superconducting state of (110)-oriented \YBCO\ films,
  expected near the surface in this geometry.
  Zero field and weak transverse field measurements
  performed using the low-energy muon spin rotation  technique
  with muons implanted few nm inside optimally-doped \YBCO-(110) films show no appearance of
  spontaneous magnetic fields below
  the superconducting temperature down to 2.9 K. Our results give an upper
  limit of $\sim$0.02 mT for putative spontaneous internal fields.
\end{abstract}
\pacs{11.30.Qc, 73.22.Gk, 74.20.Rp, 74.25.Ha, 74.72.-h}
\maketitle

In unconventional superconductors,  in addition to the global
$U(1)$ phase rotation symmetry, other symmetries
may be broken, leading to novel properties and multiple superconducting phases
\cite{review}. A particularly interesting case is broken time reversal symmetry (BTRS).
A superconducting BTRS phase has been identified
in few systems, such as Sr$_2$RuO$_4$ \cite{lukeNt98}, PrOs$_4$Sb$_{12}$ \cite{aokiPRL03},
LaNiC$_2$ \cite{hillierPRL09}, and recently in LaNiGa$_2$ \cite{hillierPRL12}.
A direct manifestation of BTRS is the appearance of
spontaneous weak magnetic fields \cite{sigristPTP98},
detected in these systems by zero field muon spin relaxation
(ZF-\msr) \cite{lukeNt98,aokiPRL03,hillierPRL09}.
 In high-$T_c$ superconductors, the non-degenerate spin-singlet $d_{\rm x^2-y^2}$-wave ($d$-wave)
order parameter preserves time reversal symmetry; thus ZF-\msr\ and other
techniques found no evidence of  bulk BTRS \cite{kieflPRL90,macDougalPRL08,SonierPRB02}.

The bulk $d$-wave order parameter, however, can be suppressed near surfaces, defects and boundaries,
where a sub-component order parameter can be stabilized leading in some cases
to a BTRS state \cite{sigristPTP98}. Whereas the best conditions for the appearance 
of a detectable effect at defects or twin boundaries in \YBCO\ (YBCO)
are not well  clarified \cite{SigristPRL95,ZhuPRB99}, 
BTRS is predicted to occur near a surface perpendicular to the 2-dimensional
${\rm CuO}_2$ planes, where the elastic scattering  of the quasiparticles experience a sign change
of the pair potential along their trajectory in the CuO$_2$ planes \cite{huPRL94}.
This process (maximal in (110) orientation) leads  to the formation of zero energy, current
carrying surface Andreev  bound states (ABS). These appear, in the quasi-particle density
of states (DOS), as a zero bias conductance peak (ZBCP) \cite{deutcherRMP05}.
The scattering leads to breaking of the $d$-wave Cooper pairs
within a few coherence lengths from the interface, giving way to
a sub-dominant pairing interaction of different symmetry (e.g.
$s$-wave) and a BTRS complex order parameter such as $d$+i$s$
\cite{sigristPTP98,fogelstromPRL97}. Emergence of a BTRS state at the surface is associated
with a spontaneous magnetic field, and leads to a shift in the energies of the ABS
in zero field, thus splitting the ZBCP.

The experimental evidence of BTRS beneath the surface of YBCO
is still controversial. ZBCP splitting has been measured in some tunneling experiments
at low temperatures ($\lesssim$10 K)  \cite{covingtonPRL97,daganPRL01,elhalelPRL07},
is absent in others indicating a predominant $d$-wave symmetry  \cite{weiPRL98},
or observed in unexpected orientations
like (100) \cite{krupkePRL99}.  Experiments carried out to detect directly the
spontaneous magnetism have also led to contradicting results.
Weak fields have been detected  near the edges of thin $c$-axis
YBCO films (that mimicks a (110)-surface) below  \Tc\ using SQUID magnetometry
\cite{carmiNt00}. Spontaneous magnetic flux was also measured near asymmetric
45$^{\rm o}$  grain boundary of $c$-axis YBCO films in zero field \cite{mannhartPRL96}.
However, fractional flux quanta were observed by  scanning  SQUID microscopy
near $c$-axis films but not in a BTRS-supporting (103) surface \cite{tafuriPRB00},
while phase sensitive measurements showed no evidence for a BTRS state
\cite{neilsPRL02}. A \bnmr\ experiment measured the field distribution
in a silver overlayer deposited on YBCO films and found
weak linewidth broadening below the \Tc\ of YBCO \cite{saadaouiPRB11}. However,
the broadening is observed in different orientations,  including $c$-axis,
pointing to a different origin than BTRS.

These conflicting results call for a direct experiment to locally probe  the magnetism
associated with a possible BTRS state near a (110) surface.  The \msr\ technique has
been the most sensitive local probe to study BTRS
appearing below $T_c$ under zero field conditions in all candidate superconductors
\cite{lukeNt98,aokiPRL03,hillierPRL09}. Weak spontaneous magnetic fields
can be detected by implanting spin-polarized
muons in the bulk and monitoring their spin depolarization
which is very sensitive to the surrounding local field.
A recently developed technique, low-energy muon spin rotation (\lem),
extends this capability to thin films, near surface region
and interfaces using  muons of tunable keV energy, stopping at
depths from a few nm to a few hundred nm \cite{morenzoni04}.
In this paper, we report ZF and weak transverse field (TF) measurements using \lem,
to directly measure the local magnetic field beneath  the surface of YBCO-(110) films.
We find no evidence of magnetism at depths from few nm to $\sim$60 nm
below the surface of YBCO, the region where the spontaneous magnetism is
expected in a BTRS scenario. We estimate an upper limit of 0.02 mT of the putative BTRS fields.

The measurements were performed at the $\mu E4$ beam-line 
of the Swiss Muon Source \cite{prokschaNIM08},
at the Paul Scherrer Institute, in Switzerland. A high-energy beam of surface muons
is moderated by solid Ar, emitting low-energy muons of few eV. These muons are
accelerated to 14 keV and directed to the sample plate.
The samples are mounted onto a plate which is electrically isolated from the cold
finger of the cryostat, and biased
to a high voltage ranging between $-12.5$ to $12.5$ kV. This allows to change
the implantation energy of the muons between $1.5$ to $26.5$ keV.   The LE-$\mu$SR
measurements are performed in a temperature range from 2.9 to 150 K. In ZF,
the stray magnetic fields at the sample position are less than 0.002 mT
in all directions. In weak TF, the field is applied perpendicular to
both the initial muon spin polarization and beam direction,
but parallel to the face of the sample.
The time evolution of the polarization of the muon ensemble implanted in the sample,
${P}(t)$, measured via detection of the emitted decay positron intensity as a function of
time after implantation, is very sensitive to the local magnetic environment.

\begin{figure}[t]
 \includegraphics[width=4cm]{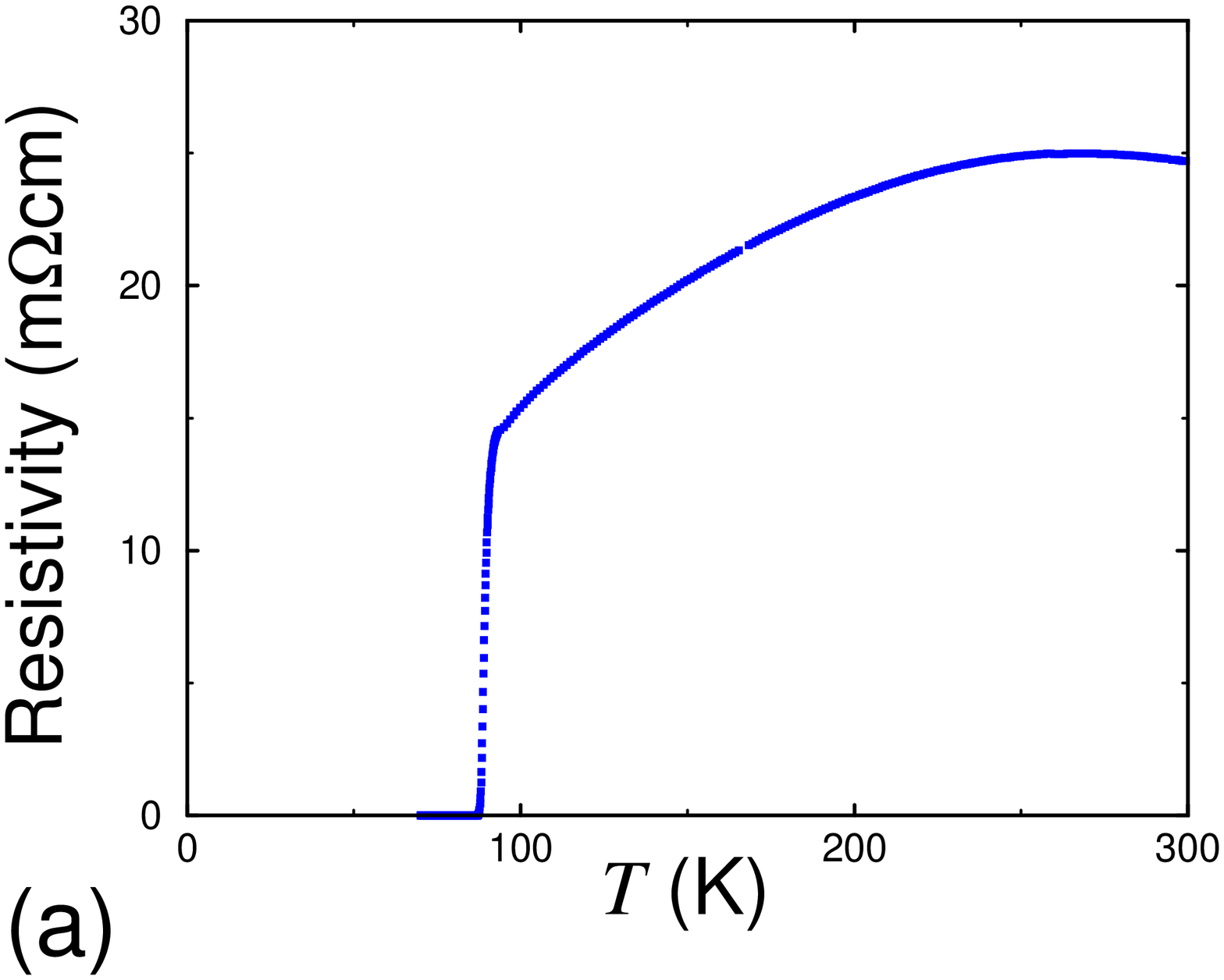}(b)
 \includegraphics[width=4cm]{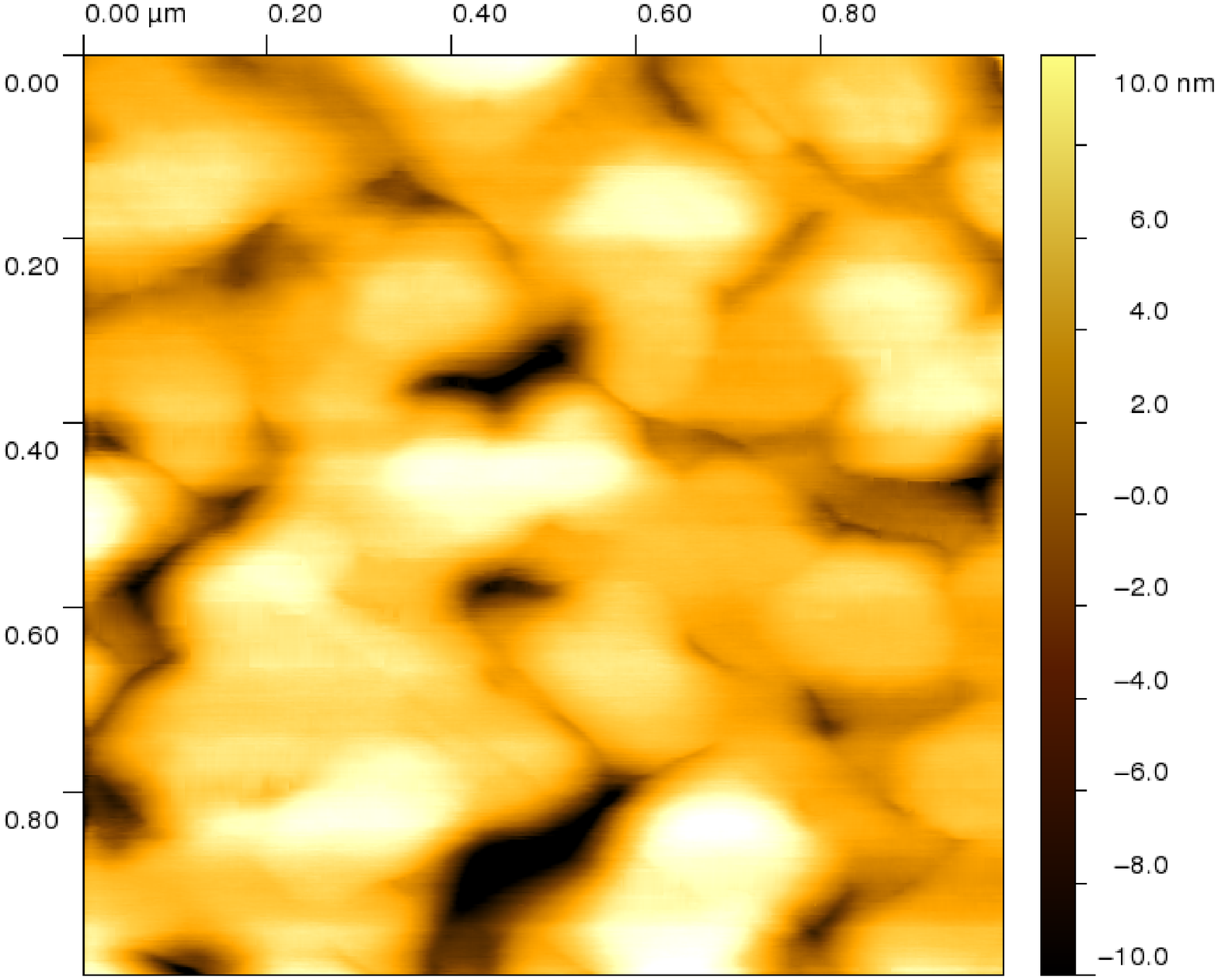}
 \includegraphics[width=3.8cm]{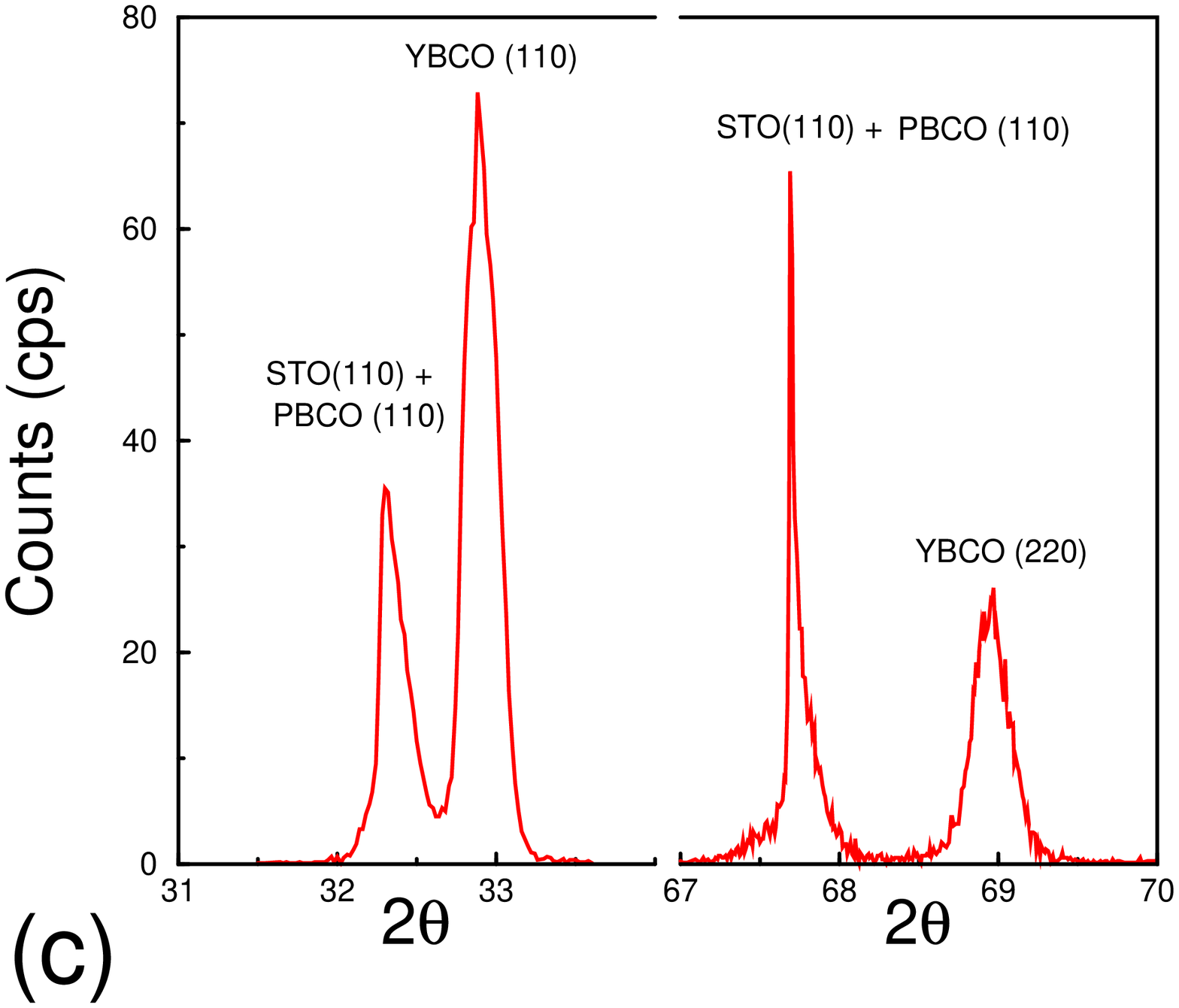} (d)
 \includegraphics[width=3.6cm]{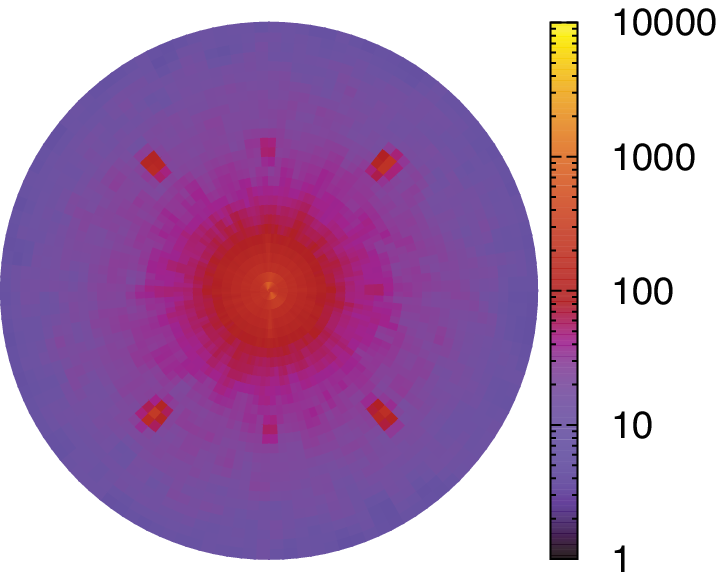}
  \caption{Typical characterization curves of the (110)-oriented films:
    (a) Resistivity along $c$-axis.
    (b) An AFM image of 1$\times$1 $\mu$m$^2$ area with an RMS roughness of about 3 nm.
     XRD Bragg peaks (c) and pole figure (d) confirming the
    (110) orientation of the films and absence of other phases.}
  \label{fig-char}
\end{figure}

The YBCO-(110) films were grown by pulsed laser deposition on (110)
SrTiO$_3$ (STO) single crystal substrates measuring $5\times5$ mm,
with a PrBa$_{2}$Cu$_{3}$O$_{7-x}$ (PBCO) template layer. The films
are optimally doped with a transition temperature $T_c=88.5$ (1.0) K,
as shown by resistivity measurements (Fig. \ref{fig-char}-(a)), 
whose temperature dependence is typical
of $c$-axis resistivity \cite{krupkePRL99}.
AFM shows a typical RMS surface roughness of less than 3 nm over an area of a few square microns (Fig. \ref{fig-char}-(b)), 
but the films are atomically flat (RMS roughness $\leq$ 1 nm) on regions of order 0.1 $\mu$m 
and covering about 80$\%$ of the surface. The overall RMS roughness of 3 nm is due to small 
regions in between with vertical steps (nanocracks) of typical depth of 10-20 nm. 
In this region the TRSB effect and associated magnetic field are expected to be smaller 
but not to disappear completely unless the orientation is (100), (010) or (001) \cite{fogelstromPRL97}. 
Moreover, muons stopping in these nanocracks are still sensitive to the fields generated in the nearby flat parts of the surface.
X-ray diffraction (XRD) confirmed the crystallinity and single
phase (110) of the films with less than 1$\%$ of other  phases (Figs. \ref{fig-char}-(c),(d)).
Rutherford back scattering (RBS) was used to check the composition and
thickness of the films and estimated the thickness of
the YBCO films of $\sim 200$ nm. A mosaic of YBCO samples of a total area of 150 mm$^2$
was mounted onto the sample holder with good electric contact. 
Monte Carlo simulations using TRIM.SP \cite{trim.sp} show that
the muon average implantation depth in YBCO ranges
from 5  to 120 nm with a corresponding range straggling of 5 and 23 nm,
for implantation energies from 1 to 25 keV, respectively.

\begin{figure}
  \includegraphics[width=\columnwidth]{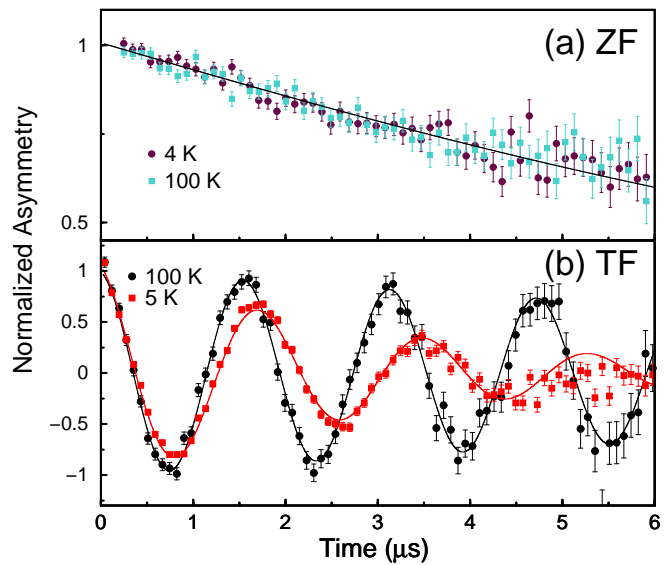}
  \caption{(a) Asymmetry in ZF at energy 2 keV, and temperature 100 and 4 K.
    (b) Asymmetry in weak TF of 4.64 mT (applied after ZF cooling)
    with muons of energy 14.2 keV at 100 and 5 K.
    Solid lines are fits explained in the text, and done using the musrfit package \cite{SuterPC12}. }
  \label{fig-asy}
\end{figure}

Typical ZF and TF \msr\ spectra are presented in Fig. \ref{fig-asy}, above and below $T_c$.
We start by discussing the TF data before focusing on the ZF results. The
TF measurements, in a field of 4.64 mT applied parallel to the surface,
allow one to characterize the Meissner state by measuring the depth profile of
the local field below the surface. The field is applied either parallel
to the $c$-axis of YBCO, so that the shielding supercurrents flow in the $ab$-planes
and one measures the in-plane London penetration depth $\lambda_{ab}$,
or by rotating the sample by 90$^{\rm o}$ where the field is perpendicular to the $c$-axis
with the shielding supercurrents flow parallel to the $c$-axis and one measures $\lambda_{c}$.
The local field $B$ is obtained from the depth dependent Larmor spin precession frequency
$\omega=\gamma_{\mu} B$, where $\gamma_{\mu}/2\pi=135.5$ MHz/T
is the $\mu^+$ gyromagnetic ratio. The samples in TF measurements are mounted on nickel plated
sample holder, where the muons missing the sample depolarize very quickly
(within few ns) and do not contribute to the polarization signal \cite{kieflPRB10,saadaouiPC12}.

\begin{figure}
  \includegraphics[width=\columnwidth]{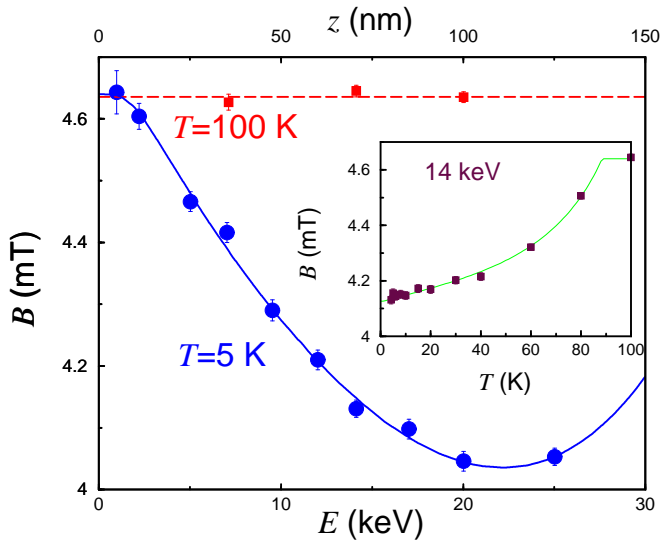}
  \caption{The energy dependence of the average magnetic field
    in the Meissner state at 5 K.
    The applied field of 4.64 mT as measured  in the normal state
    at 100 K (dashed line), is applied parallel to the $c$-axis
    direction of YBCO films. The  average depth corresponding to
    the muons energy as simulated by TRIM.SP is displayed
    on the top axis. Inset: variation of the average magnetic field  versus temperature
    with 14 keV muons stopping at an average depth of 65 nm. The solid line
    is a fit using a $d$-wave gap function \cite{prozorovSST06}.}
  \label{fig-wtf}
\end{figure}

Upon zero-field cooling below \Tc\ and then applying the TF, we see a significant
reduction in the precession frequency as expected in the Meissner state \cite{kieflPRB10}.
The mean internal field as a function of the beam energy at 5 K is plotted
in Fig. \ref{fig-wtf} for an external field applied parallel to the $c$-axis.
A fit using a cosine hyperbolic function, valid for thin films \cite{JacksonPRL00},
yields $\lambda_{ab}\sim 160$(10) nm.
In a field perpendicular to the c-axis, we obtain $\lambda_{c}\sim 450(20)$ nm,
thus confirming the overall (110) orientation of the films.
The temperature dependence of the internal field is plotted in the inset
of Fig. \ref{fig-wtf}.
The data are consistent with a single $d$-wave order parameter
in the investigated region \cite{prozorovSST06}.
From these weak-TF measurements,
we can also rule out the appearance of strong spontaneous fields
near the surface  since in both orientation and throughout
the film we observed only the diamagnetic shift characteristic of the Meissner state.

\begin{figure}
   \includegraphics[width=\columnwidth]{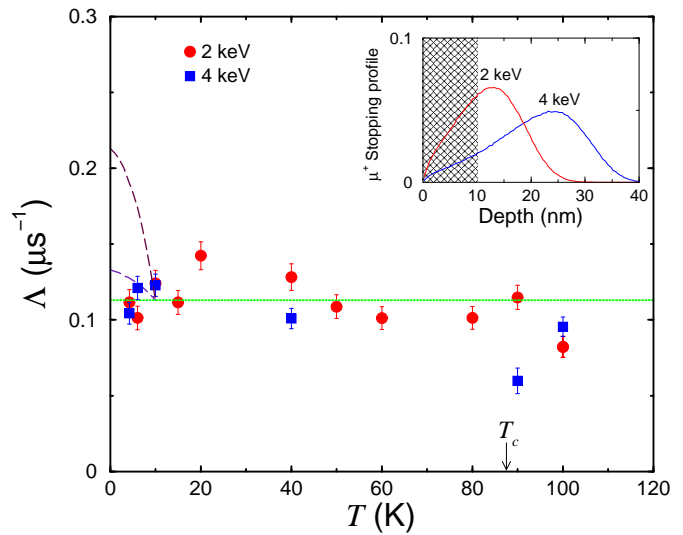}
   \caption{The electronic depolarization rate versus temperature
     with slow muons of energy 2  and 4 keV implanted in YBCO.
     The dashed lines are guidelines representing the effect of BTRS fields of width 0.02 and 0.1 mT
     on the ZF depolarization rate. Inset shows the stopping profile of the implanted muons, where the
     average RMS depth for 2 and 4 keV muons is 12(5) and 21(8) nm, respectively.
     Dashed region corresponds to a region where BTRS is expected.}
   \label{fig-zf}
\end{figure}

In bulk superconductors the most sensitive signature of spontaneous magnetism due to BTRS
is a characteristic monotonic increase of ZF depolarization rates setting in below
the critical temperature of the phase transition with BTRS.
The ZF asymmetry in (110)-oriented YBCO films as presented in Fig. \ref{fig-asy}-(a) shows
no apparent difference in cooling below \Tc. Beam energies of 2 and 4 keV are chosen
to maximize the number of probes stopping 
in the region where TRSB is expected \cite{fogelstromPRL97}.
The asymmetry decays with a small depolarization rate ruling out the presence of large
internal fields. If a long-range ordered magnetic state is present, the \msr\
spectra should display precession at a frequency set by the average magnetic field
at the muon site. In a disordered magnetic field state, the spin polarization decays faster
with a rate set by the overall field width.
We performed measurements of the ZF-\msr\ asymmetry from 120 down to 2.9 K, and fitted
the data to the function \cite{lukeNt98,aokiPRL03,hillierPRL09}
\begin{equation}
  G(t) = A\Big[\frac{1}{3}+\frac{2}{3}(1-\Delta^2 t^2)e^{-\frac{1}{2}\Delta^2 t^2}\Big]
  e^{-\Lambda t} + A_{\rm bg}e^{-\lambda_{\rm bg}t},
\end{equation}
where the first term accounts for YBCO, and the second for
the temperature independent background signal from the muons missing the sample
and hitting the Ag-coated sample holder used in the ZF experiment \cite{saadaouiPC12}.
The observed signal is dominated by YBCO (more than 50$\%$ fraction). 
Ag has a low relaxation rate ($\sim 0.02$ $\mu$s$^{-1}$), and 
its contribution to the signal has been measured separately and included in the fit function of Eq. 1. 
We have found similar spectra when the backing signal is Nickel which relaxes 
completely the spins of the muons missing the sample.
For the ZF case we preferred to present only the data on Ag as it is a non magnetic material itself.
The depolarization in YBCO, caused by the nuclear moments
(of width $\Delta=0.08$ $\mu$s$^{-1}$), is accounted for by
the Kubo-Toyabe function, multiplied by an exponential accounting for the depolarization
of electronic origin  \cite{lukeNt98,aokiPRL03,hillierPRL09}.
The static depolarization rate, $\Lambda$, is proportional
to the width  of the local magnetic field distribution, $\Lambda \simeq \gamma_{\mu}\Delta B$.

The temperature variation of $\Lambda$ at 2 and 4 keV is plotted in Fig. \ref{fig-zf}.
This shows no noticeable increase of the depolarization rate below \Tc\
down to 2.9 K. The results are consistent with a field distribution
dominated by randomly oriented nuclear dipolar moments in YBCO.
Experiments carried out on films with double the surface roughness (not shown here) yielded
no difference with the measurements presented here, ruling out
the effect of surface quality of the films on the depolarization rate in ZF.
We can rule out the appearance of a BTRS state, as the spontaneous disordered magnetic
fields would enhance $\Lambda$ upon entering the phase supporting BTRS
as demonstrated by \msr\ in the bulk on other systems \cite{lukeNt98,aokiPRL03,hillierPRL09}.
For comparison, we plot in Fig. \ref{fig-zf} the expected behavior
of the depolarization rate for internal fields of 0.02 mT
and 0.1 mT (dashed lines), which have been typically detected in BTRS states by ZF-\msr.
From the average scatter of the data one can draw an upper limit of 0.02
$\mu{\rm s}^{-1}$ corresponding to spontaneous magnetic fields of $\sim0.02$  mT.

As mentioned above, the proposal of BTRS near (110) surfaces stems mainly from the prediction
and observation of ZBCP splitting in zero field \cite{fogelstromPRL97,covingtonPRL97,krupkePRL99}.
Fogelstr\"om \etal\ showed in Ref. \cite{fogelstromPRL97} that the scale of the field
associated with the splitting is of the  order of a
fraction of the bulk critical field, $B_0\sim \phi_0/\pi^2\xi \lambda$.
The depth of the region where spontaneous magnetism appears is not uniquely
predicted. However, it is reasonable to assume an extent of few coherence lengths
in the ab-plane $\xi \simeq$ 2 nm. At 2 and 4 keV, a sizable fraction
of muons (30 and 15\%, respectively) stop in this depth range (shadowed area, inset of Fig. \ref{fig-zf}).
Moreover, also the muons stopping outside this region should still detect the stray fields.
The width of the stray fields due to BTRS spontaneous fields
extending over length scale of $\xi$ may be approximated by
$\Delta B \sim B_{\rm s} (\xi/z)^3$, where $B_{s}$ are estimates from tunneling,
and $z$ is the muon stopping depth.  Thus at an average depth of 12 nm at 2 keV,
using an estimate of $B_s=0.1 B_0 \approx 0.1$ T, we expect
$\Delta B \sim 0.4$ mT, much higher than our finding.
Electron spin resonance performed with a 60 nm thick layer of active
spins on YBCO-(110), estimated the strength of stray fields at about 0.2 mT \cite{greeneJSC00}.
A similar experiment performed by \bnmr\ with polarized \Li\ spins implanted in an Ag overlayer on
YBCO-(110) draw an upper limit of BTRS at 0.02 mT \cite{saadaouiPRB11}.
One has to notice that these experiments are done outside the superconductor, thus one
expects a reduced sensitivity with respect to the present experiment.

The preferred muon site in \YBCO\ (close to the chain oxygen) and all possible secondary sites
do not have lattice symmetry, so that a cancellation of fields from supercurrent
with lattice symmetry can be excluded \cite{muonsite}.
Moreover, our experiment corresponds to the regime of zero transmission interface
which significantly enhances BTRS and the corresponding ZBCP splitting \cite{tanumaPRB01},
and is free of complication of possible unusual  proximity effects \cite{kohenPRL03}.
These together with the fact that the measurements are performed directly in the YBCO
make the present experiment particularly sensitive to magnetism.
The absence of spontaneous BTRS fields in our experiment is in agreement with
tunneling measurements where the ZBCP splitting was only observed in an applied field
and vanishes in zero field \cite{elhalelPRL07,weiPRL98,apriliPRL99,beckPRB04,yehPRL01}.

In conclusion, we have conducted a depth-resolved \lem\ study of the field
distribution near the surface of YBCO films to look for direct evidence
of spontaneous magnetism associated with a possible time reversal symmetry
complex order parameter that develops near a (110) surface.
We detected no spontaneous internal fields in YBCO, and consequently
we established an upper limit of about 0.02 mT for BTRS fields at low temperature.
This field value is much weaker than that expected by tunneling experiments
and various theories, where the ZBCP splitting is associated with much stronger spontaneous fields.

We  thank M. Horisberger for depositing Ni and Ag on sample plates,
M. D$\ddot{\rm o}$beli for performing RBS measurements, S. Rolf for assisting with AFM
measurements, B. M. Wojek for helping with fitting, and R. F. Kiefl
and M. Sigrist for useful discussion.
The authors HS and EM acknowledge the financial support of the MANEP program, and HH and
PP, of the Wihuri Foundation.


\end{document}